\documentclass[prl,aps,twocolumn,floats,nofootinbib]{revtex4}
\usepackage{amsmath,amssymb,graphicx,psfrag}

\begin{document}
\renewcommand{\ni}{{\noindent}}
\newcommand{\dprime}{{\prime\prime}}
\newcommand{\be}{\begin{equation}}
\newcommand{\ee}{\end{equation}}
\newcommand{\bea}{\begin{eqnarray}} 
\newcommand{\eea}{\end{eqnarray}}
\newcommand{\la}{\langle}
\newcommand{\ra}{\rangle} 
\newcommand{\dg}{\dagger}
\newcommand\lbs{\left[}
\newcommand\rbs{\right]}
\newcommand\lbr{\left(}
\newcommand\rbr{\right)}
\newcommand\f{\frac}
\newcommand\e{\epsilon}
\newcommand\ua{\uparrow}
\newcommand\da{\downarrow}
\title{Can many-body localization persist in the presence of long-range interactions or long-range hopping?}
\author{Sabyasachi Nag and Arti Garg}
\affiliation{Condensed Matter Physics Division, Saha Institute of Nuclear Physics, 1/AF Bidhannagar, Kolkata 700 064, India}
\vspace{0.2cm}
\begin{abstract}
\vspace{0.3cm}
We study many-body localization (MBL) in a one-dimensional system of spinless fermions with a deterministic aperiodic potential in the presence of long-range interactions or long-range hopping. Based on perturbative arguments there is a common belief that MBL can exist only in systems with short-range interactions and short-range hopping. We analyze effects of power-law interactions and power-law hopping, separately, on a system which has all the single particle states localized in the absence of interactions.  Since delocalization is driven by proliferation of resonances in the Fock space, we mapped this model to an effective Anderson model on a complex graph in the Fock space, and calculated the probability distribution of the number of resonances up to third order. Though the most-probable value of the number of resonances diverge for the system with long-range hopping ($t(r) \sim t_0/r^\alpha$ with $\alpha < 2$), there is no enhancement of the number of resonances as the range of power-law interactions increases.  This indicates that the long-range hopping delocalizes the many-body localized system but in contrast to this, there is no signature of delocalization in the presence of long-range interactions. We further provide support in favor of this analysis based on dynamics of the system after a quench starting from a charge density wave ordered state, level spacing statistics, return probability, participation ratio and Shannon entropy in the Fock space. We demonstrate that MBL persists in the presence of long-range interactions though long-range hopping with $1<\alpha <2$ delocalizes the system partially, with almost all the states extended for $\alpha \le 1$. {\it Even in a system which has single-particle mobility edges in the non-interacting limit, turning on long-range interactions does not cause delocalization.} 
 
\vspace{0.cm}
\end{abstract} 
\maketitle
\section{I. Introduction}

 Many-body localization (MBL) has been a topic of intense research in condensed matter physics, both theoretically~\cite{Anderson,Altshuler,Basko,Gornyi,Huse2007,Huse2010,Huse2013,Abanin,Mueller,Huse2014,Bardarson,Laumann,Bera,Alet,Sdsarma,Pal,Subroto,shastry,Huse,Altman,ME1,garg,Bera2,Baldwin,sudip,abanin_rev,sdsarma2018,Achim,Ehud,sdsarma2019} and experimentally~\cite{expt,expt2,InO,optical,ions,diamond}. MBL, which is realized in quantum systems that are sufficiently disordered and interacting, is an interesting and unusual phase of matter in many aspects. An isolated quantum system in the MBL phase is non-ergodic and hence, challenges the basic foundations of quantum statistical physics~\cite{Huse,Altman}.
Local observables in the MBL phase do not thermalize leading to the violation of eigenstate thermalization hypothesis (ETH)~\cite{Deutsch,Srednicki,Rigol}. This results in a rich behaviour of entanglement entropy~\cite{Nayak,Huse2013, Sdsarma} and a long time memory of the initial state in local observables. Infinite temperature MBL phase has been shown to have an extensive number of local integrals of motion~\cite{Abanin,Mueller} and hence is similar to integrable systems~\cite{Huse2014,shastry}. MBL can prevent heating in a periodically driven system~\cite{MBL_periodic} and can also help in stabilizing new phases of matter like time crystal~\cite{ions,diamond}.
In short, exploration of MBL phase has revealed a plethora of exotic physics.

 It is natural to ask, in what kind of systems MBL can be realized. Majority of theoretical studies have focused on MBL in 1-dimensional systems in the presence of short-range interactions which includes extensive numerical studies using state-of-the-art exact diagonalization~\cite{Huse2007,Huse2010,Huse2013,Bardarson,Bera,Alet,Sdsarma,Pal,Subroto,Huse,ME1,garg,Bera2,
Baldwin,sdsarma2019},  and analytical approaches like renormalization group and approximate locator expansion in the Fock space~\cite{RG,logan}. In his seminal work~\cite{Anderson}, Anderson showed that single particle localization can not occur in the presence of long-range hopping $t \sim 1/r^\alpha$ for $\alpha \le d$ where $d$ is the dimension of the system.  Generalization of Anderson's criterion for a small subset of interacting particles showed that MBL can not survive in systems with power-law interactions for $\alpha < 2d$~\cite{Burin,Dipole}. Most of the numerical attempts to establish MBL in systems with long-range interactions in one-dimension have lead to a consensus that MBL can not survive in systems with power-law interactions and power-law hopping with $\alpha <2$~\cite{Burin,Dipole,Mirlin,Sarma2015} though for $\alpha >2$ MBL occurs~\cite{Dipole,Sarma2015}. On the other hand, perturbative treatment of effective Anderson model in the Hilbert space for disordered spin chain with power-law interactions showed existence of MBL for $\alpha <1$~\cite{Heyl}. Thus, effect of long-range interactions on MBL phase is an issue of debate in the community~\cite{note_lr}. 

On experimental side, there are many real systems like trapped ions, defects in solid-state systems and ultracold polar molecules~\cite{ions2,Rydberg,defects,polar}, which are examples of interacting disordered systems and are hence ideal for investigating MBL, but show power-law decay of interactions with distance. Hence, it is crucial to understand the physics of many-body-localization in systems with long-range interactions and long-range hopping.
Interestingly, recent experiments~\cite{ions,diamond} with dipoles have established a time-crystal phase, which is supposed to be stabilized due to the existence of MBL-like phase, in a system where the above mentioned argument would suggest that MBL cannot arise. With this motivation, the question we want to answer in this manuscript is whether MBL can survive, at all, in systems with long-range interactions or long-range hopping.

In this work, we study a model of spinless fermions in one dimension in the presence of an aperiodic deterministic potential~\cite{Fishman,Sarma1990,Sarma_nonint,garg}, which can also be viewed as a generalized Aubry-Andre potential~\cite{AA,Subroto}. The virtue of this model over a fully random disorder potential is that by changing parameters in this model, one can realize a system with all single particle states localized/delocalized as well as the system that has single particle mobility edges. Unlike most of the earlier works~\cite{Burin,Dipole,Mirlin}, we study the effect of power-law interactions and power-law hopping separately in this work. For the system with power-law interactions, only nearest-neighbour hopping is considered while for the case of power-law hopping the system has nearest-neighbour interactions. 

The main findings of our work are summarized below. We found that the long-range interactions have very weak effect on MBL not only on systems which have all single particle states localized but also for systems that have single particle mobility edges in the non-interacting limit. In contrast to this, long-range hopping delocalizes the system. 
By mapping the model mentioned above onto an effective Anderson model in the Fock space, we calculated the probability distribution for number of resonances in the Fock space up to third order. In contrast to resonance arguments for a small subset of interacting particle, our calculation, which captures correlation between all particles in the system, shows that the most probable value of the number of resonances increases for the power-law hopping, ($t\sim t_0/r^\alpha$), as $\alpha$ decreases, diverging in the thermodynamic limit for $\alpha <2$; though there is no change in the most-probable value of the number of resonances for the system with power-law interactions for all $0.5\le \alpha \le 3$ values considered. Since many-body delocalization is driven by proliferation of resonances in the Fock space, our resonance count analysis indicates that the delocalization should get enhanced as the range of hopping increases though the long-range interactions have very weak effect on the number of resonances in the Fock space.
  
 Furthermore, we study dynamics of the system after a quench starting from a charge density wave (CDW) ordered state. For the system where all the many-body states are localized in the presence of nearest-neighbour interactions and hopping, turning on power-law hopping helps the system in relaxing resulting in reduced value of the density imbalance in the long time limit which is accompanied by enhanced growth of the entanglement entropy for smaller $\alpha$ values. In the presence of power-law interactions, on the other hand, no significant change is observed in the density imbalance or the entanglement entropy, for various values of $\alpha$ studied. We also see clear signatures of enhanced delocalization in the presence of power-law hopping for $\alpha < 2$ in the return probability. The return probability has a faster power-law decay for longer range hopping which also makes its saturation value smaller as $\alpha$ decreases, though the return probability in the long time limit remains same for all ranges of interactions studied which indicates that the system retains memory of the initial state even in the presence of long-range interactions.

 In the end, we study static properties of the system like level spacing statistics, participation ratio and Shannon entropy in the Fock space and provided an energy resolved phase diagram as a function of $\alpha$. In the presence of power-law interactions, the ratio of consecutive level-spacings remains close to the value expected for Poissonian distribution for all the energy eigenstates, in the entire range of $\alpha$ considered, clearly indicating that all the many-body states remain localized even for long-range interactions with $\alpha <2$. In contrast to this, as the range of power-law hopping increases, the fraction of many-body states for which the level spacing ratio coincides with Wigner-Dyson value increases. For power-law hopping with $1< \alpha <2$ a large fraction of states have level spacing ratio equal to the Wigner-Dyson value and for $\alpha \le 1$, the entire spectrum gets delocalized. Similar conclusion is drawn from the analysis of the participation ratio in the Fock space.

 Based on the above analysis, we demonstrate that MBL can persist in the presence of long-range interactions for system which has all single particle states localized. A natural question that emerges and needs to be addressed is what happens if the non-interacting system has single particle mobility edges(SPME).  We study dynamics of the system after a quench from the charge density wave ordered state in this situation as well. Quite surprisingly, even in this scenario, the long-range interactions do not cause any delocalization and the system continues to show many-body mobility edges. 

The rest of the paper is organized in the following manner. In Section II, we introduce the model explored in this work and set the notations for the rest of the paper. In section III, we map this model to an effective Anderson model in the Fock space and calculate the number of resonances in the Fock space up to third order. In section IV we discuss results for the system where all the single particle states are localized. Section IV-A provide the details of level spacing statistics. In section IV-B, we describe the dynamics of the system after a quench starting from a charge density wave ordered state. Section IV-C presents results for the return probability. In the end, in section IV-D, we obtain the energy resolved phase diagram based on level spacing statistics, participation ratio and Shannon entropy.

 Based on all this analysis, we demonstrate the surprising result that MBL can persist in the presence of long-range interactions though long-range hopping causes delocalization. The question that comes naturally is : are interactions themselves localizing the system. To rule this out, in the end of section IV, we explore effects of long-range interactions in the parameter regime of the model under consideration where all the single particle states are delocalized and show that the system remains fully extended even in the presence of long-range interactions. In section V, we study effect of power-law interactions on the system which has single particle mobility edges. Finally we summarize our results and conclude with some remarks on connected works in section VI. 

 \section{II. Model}
We study a one-dimensional model of spin-less fermions which is described by the Hamiltonian of the form 
\bea
H=-t_0\sum_{i}[c^\dagger_ic_{i+1}+h.c.] + \sum_i h_i n_i + V\sum_i n_in_{i+1} \nonumber \\
+\sum_{j>i+1}\lbr\f{V_1}{(j-i)^\alpha}n_in_j - \f{t_1}{(j-i)^\alpha}[c^\dagger_ic_j+h.c.]\rbr
\label{model}
\eea
Here $t_0$ is the nearest neighbor hopping amplitude on a one dimensional chain with open boundary conditions, $V$ is the nearest neighbor repulsion and $h_i$ is the on site potential of form $h_i=h \cos(2\pi\alpha i^n+\phi)$ where $\alpha$ is an irrational number and $\phi$ is an offset~\cite{Fishman,Sarma1990,Sarma_nonint}. 
First consider the non-interacting limit of the model in Eq.(\ref{model}) in the presence of only nearest neighbour hopping (that is $V=V_1=t_1=0$). In this limit for $n=1$, which is also known as the Aubry-Andre (AA) model~\cite{AA}, all the single particle states are delocalized for $h < 2t$. In this work we mainly study this model for $n < 1$ for which the system has single particle MEs at $E_c=\pm|2t-h|$ for $h < 2t$~\cite{Sarma1990}. For $h>2t$, all the single particle states are localized for any value of $n$.
$V_1$ and $t_1$ are the coefficients of power-law interaction and power-law hopping respectively. 

For the system with nearest-neighbour hopping and repulsion between fermions ($V_1=t_1=0$ in Eq.(\ref{model})), this model has been studied using exact diagonalization~\cite{Subroto,garg}. In earlier work~\cite{garg}, we demonstrated that for $h < 2t$, where the non-interacting system has single particle mobility edges, the interacting system with nearest-neighbour (NN) interactions ($V_1=t_1=0$) shows MBL only if the chemical potential does not lie between the two single particle mobility edges. Though for $h>2t$, the NN interacting system can show MBL at any filling. In both the cases, the many-body states at the top and the bottom of the spectrum are localized and the width of the delocalized regime increases as the strength of $V$ increases in comparison to $h$. For very strong disorder $h\gg 2t$, the NN interacting system shows an infinite temperature MBL phase where all the many-body states are localized.

In this work we either consider the case $V_1=V$ and $t_1=0$ which corresponds to the system with power-law interactions between fermions or we consider the system with $V_1=0$ and $t_1=t_0=1$, which represents the case of power-law hopping. Below we develop the resonance argument by mapping this model to an effective Anderson model in the many-body Fock space.  
\begin{figure}[h!]
\begin{center}
\hspace{1cm}
\includegraphics[width=2.35in,angle=-90]{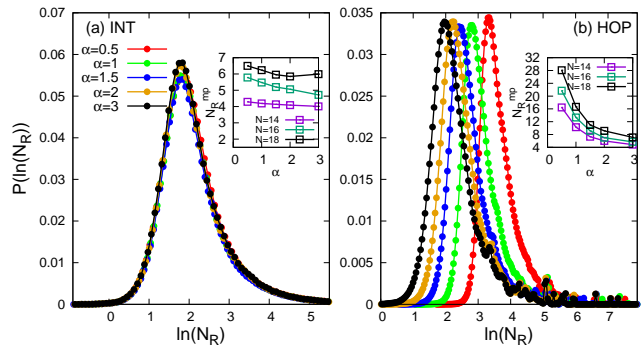}
\caption{The left panel shows probability distribution $P(\ln(N_R))$ for the system with power-law interactions and the right panel shows the results for the system with power-law hopping for various powers $\alpha$. Peak position of the distribution shifts to larger $N_R$ values for longer-range hopping but the peak position remains more or less the same for various values of $\alpha$ for the system with power-law interactions. The distributions are shown for $L=18$, for the half-filled system with $h=5t_0$ and $V=1.0t_0$, averaged over 500 independent configurations. The insets show most probable value of the number of resonances $N_R^{mp}$ vs $\alpha$ for various system sizes, for configuration averaging from 5000 to 500 for $L=14$ to $L=18$.}
\label{prob_res}
\vskip-1cm
\end{center}
\end{figure}

\section{III. Resonances in many-body Fock space}
We first revise the resonance count argument for the non-interacting Anderson model in the presence of power-law hopping.
For the non-interacting model with power-law hopping, resonant pairs of state are those for which $|h_i-h_j|\le t_0/|r_j-r_i|^\alpha $. The expected number of resonant states at a distance $R$ from a central state is 
\be
N_{1}(R) \sim (\rho R^d) \frac{t_0/R^\alpha}{2h}
\label{N1}
\ee
Here $\rho$ is the density of particles. For $\alpha < d$, $N_1(R)$ diverges as $R \rightarrow \infty$, which implies that any single particle state resonates with any other state at arbitrary distance and hence localization is impossible. 

To develop a similar argument for the interacting many-body system in the presence of an aperiodic potential, we map the Hamiltonian in Eq.~\ref{model}  to an effective Anderson model defined in the Fock space of spinless fermions which has $^LC_N$ configurations for $N$ fermions on $L$ sites chain.  The basis states are specified by the occupancies of each site $|{n_i}\rangle$ where $n_i$ is 1(0) if the site $i$ in real space is occupied (unoccupied). Let us first consider the case of power-law interactions. The effective Hamiltonian in the Fock space basis has the following form:
\be
H_{eff} = \sum_l \epsilon_l |l\ra \la l|+ \sum_{lm} \hat{T}_{lm}|l\ra \la m|
\label{heff}
\ee 
with 
\bea
\epsilon_l = \sum_i h_i \la l|n_i|l \ra +V\sum_{j>i}\f{\la l|n_in_j|l \ra}{(j-i)^\alpha}\\ \nonumber
\hat{T}_{lm} = -t_0\sum_i \la l|c^\dagger_ic_{i+1}+h.c.|m \ra
\label{matel}
\eea
 with ${|l\ra}$ representing configurations in the Fock space. Note that the mapping from Eq.(\ref{model}) to Eq.(\ref{heff}) is exact and has been used earlier in many works~\cite{Heyl,map}. The local energy $\epsilon_l$ has contribution from  interactions among all the particles in the system. Although the effective Hamiltonian is non-interacting, the problem is still hard to solve due to complexity of the graph over which the model is defined. Starting from a configuration in the Fock space, system can hop onto many different configurations, which are of order $L$ for a $L$ site system and the connectivity of the graph varies from configuration to configuration. 

In analogy to single particle Anderson model, delocalization here is also driven by proliferation of resonances in the Fock space. 
Again identifying the resonant pair of states as those for which $|\epsilon_l-\epsilon_m| \le |\hat{T}_{lm}|$, the number of resonant pairs, starting from $l$th configuration is
\be
N_R(l) = \sum_{m=1}^{V_f} \frac{|\hat{T}_{lm}|}{|\epsilon_l-\epsilon_m|}
\label{NR}
\ee 
The sum over $m$ runs over all the $V_f$ number of configurations in the Fock space such that $\epsilon_l \ne \epsilon_m$. Note that $N_R$ gives the number of resonances up to first order in $t_0/W$ where $W$ is the variance of $\epsilon_l$. 

In order to calculate $N_R(l)$ numerically for every basis state $|l\ra$ for a given realization of the aperiodic potential, we first calculate the matrix elements of the effective Anderson model in Eq. (\ref{matel}), namely $\epsilon_l$ and $\hat{T}_{lm}$. After calculating $N_R(l)$ for all allowed values of $l$ in the Fock space, we obtain the probability distribution of $\ln(N_R)$ for a given realization of the aperiodic potential, and then average it for many independent realizations of the aperiodic potential.  Fig.~\ref{prob_res} shows the probability distribution of $\ln(N_R)$ for the half-filled system with $h=5t_0$ and $V=t_0$.
 As shown in the left panel of Fig.~\ref{prob_res}, neither the peak nor the width of the distribution changes significantly as the range of interaction increases with $\alpha$ varying from three to 0.5. The inset shows the most-probable value  $N_R^{mp}$ as a function of the range of interaction for various system sizes. There is a  weak increment seen in the value of the number of resonances with decrease in $\alpha$. 
\begin{figure}[h!]
\begin{center}
\hskip-1cm
\includegraphics[width=2.35in,angle=-90]{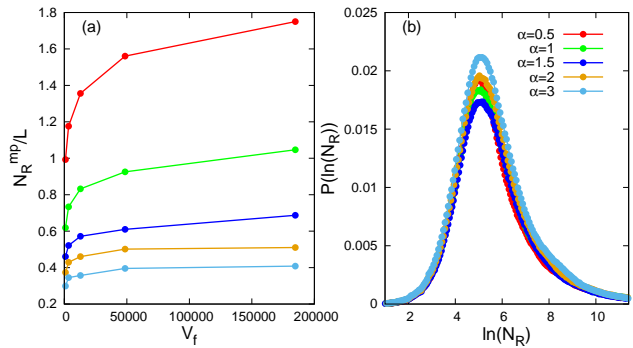}
\caption{Left Panel: $N_R^{mp}/L$ vs $V_f$, where $V_f$ is the dimension of the Fock space, for the system with power-law hopping. Though $N_R^{mp}/L$ increases with $V_f$ for small values of $V_f$ for all the cases of power-law hopping studied, for $\alpha >2$, $N_R^{mp}/L$ saturates in the limit of large $V_f$. On the other hand, for $\alpha <2$, $N_R^{mp}/L$ keeps increasing monotonically with $V_f$ even in the limit of large $V_f$ indicating diverging $N_R^{mp}/L$ in the thermodynamic limit. The data shown is averaged over (5000 to 500) configurations for $L=12$ to $20$.  Right Panel shows the probability distribution of $\log(N_R^{3rd})$ obtained within 3rd order resonance calculation for the system with power-law interactions. No significant change in the distribution or the most-probable value of $N_R^{3rd}$ is observed as the range of interaction increases. The data shown is for $L=16$ and averaged over 500 independent configurations.}
\label{mp_res}
\vskip-1cm
\end{center}
\end{figure}

Now consider the system with power-law hopping. The effective Hamiltonian in the fermionic Fock space again has the form as in Eq.(~\ref{heff}) with $\epsilon_l = \sum_i h_i \la l|n_i|l\ra +V\sum_{i}\la l|n_in_{i+1}|l \ra$ and 
\be
\hat{T}_{lm} = -t_0\sum_{j>i} \f{\la l|[c^\dagger_ic_{j}+h.c.]|m \ra}{|r_j-r_i|^\alpha}
\ee
 As shown in the right panel of Fig.~\ref{prob_res}, the most probable value of the number of resonances $N_R^{mp}$ increases with the range of hopping. Hence, one should expect enhanced many-body delocalization in the presence of long-range hopping.
In order to determine whether the number of resonances diverge in the thermodynamic limit below certain critical $\alpha$, below which MBL will not survive due to proliferation of resonances, we looked at the system size dependence of normalized number of resonance $N_{R}^{mp}/L$. The reason we chose this normalization is because connectivity of a configuration increases linearly with the system size $L$. The left panel of Fig.~\ref{mp_res} shows $N_R^{mp}/L$ vs $V_f$, where $V_f$ is the dimension of the Fock space, for various values of $\alpha$. Though for small system sizes, $N_R^{mp}/L$ increases with $V_f$ for all the cases of power-law hopping studied, for $\alpha >2$, $N_R^{mp}/L$ shows saturation in the limit of large system size and does not increase for larger $V_f$ values. On the other hand, for $\alpha <2$, $N_R^{mp}/L$ keeps increasing monotonically even in the large $V_f$ limit and hence diverges in the thermodynamic limit indicating complete delocalization of the system. Thus, the resonance count up to first order, shows that power-law hopping with $\alpha <2$ causes many-body delocalization. But long-range interactions probably have no significant effect towards delocalization of the system.

 But based on the first order calculation of $N_R$, we can not rule out delocalization in the presence of long-range interactions due to higher order resonant processes. It is important to realize that for effective Anderson model in the Fock space, it is almost impossible to calculate contribution of the infinite series, but we extended the calculation up to third order in $t_0/W$ for the case of power-law interactions. Up to third order, the number of resonances are given by $N_{R}^{3rd}$
\be
 =\sum_m \f{|\hat{T}_{lm}|}{|\epsilon_l-\epsilon_m|} \lbs 1+\sum_n^\prime\f{|\hat{T}_{mn}|}{|\epsilon_m-\epsilon_n|}\lbr 1+\sum_p^{\prime\prime}\f{|\hat{T}_{np}|}{|\epsilon_n-\epsilon_p|}\rbr \rbs
\ee 
Here $\sum_n^\prime$ represents the restricted sum over $n$ values, such that $n\ne m\ne l$ and $\sum_p^{\prime\prime}$ is the sum over $p$ such that $p\ne n,m,l$.  
The right panel of Fig.~\ref{mp_res} shows the distribution of $\log(N_R^{3rd})$ for various cases of power-law interactions. Though the most-probable value within 3rd order calculation is larger than that within first order for any $\alpha$, as expected, there is no change in the distribution of $N_R$ as the range of interaction increases. The most-probable value does not increase for longer range interactions, rather it remains more or less same for all the values of $\alpha$ studied, like in the first order calculation. 

Hence, the number of resonances in Fock space, calculated up to third order, do not show signatures of delocalization in the presence of long-range interactions ($\alpha < 2$), in contrast to the resonance argument based on a few particle resonant conditions~\cite{Dipole,Burin}. 
In the following sections, we provide results for level spacing statistics, quantum quench, Renyi entropy, and return probability, all of which show consistency with the resonance arguments presented above. This indicates that probably the number of resonances in the Fock space, with even higher order terms included, never diverge for power-law interactions.

\begin{figure}[h!]
\begin{center}
\hskip-0.5cm
\hspace{-1cm}
\includegraphics[width=2.5in,angle=-90]{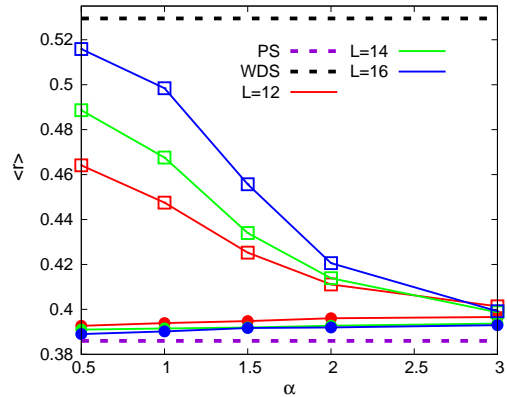}
\caption{Average level spacing ratio of successive gaps $\la r \ra$ vs $\alpha$ for various system sizes. In the presence of power-law interactions (filled circles), $\la r \ra$ does not increase either with $\alpha$ or with the system size $L$ and always remains close to the Poissonian value which indicates that MBL persists even in the presence of long-range interactions. But for the system with power-law hopping (hollow squares), $\la r \ra$ increases with decreasing $\alpha$ from Poissonian value to Wigner-Dyson value. Also for $\alpha < 2$, $\la r \ra$ shows significant increase with $L$ approaching the Wigner Dyson value in the thermodynamic limit. This shows that long-range hopping with $\alpha <2$ delocalizes the system. The data shown is averaged over (4000-200) independent configurations for $L=12$ to $L=16$.} 
\label{rn}
\vskip-1cm
\end{center}
\end{figure}
\section{IV. Effect of long-range interactions or hopping on the system with all single particle states localized}
In this section we discuss the effects of power-law interactions and hopping, separately, on the system where all the single particle states are localized. For the choice of parameters made, in the presence of nearest-neighbour interactions along with nearest neighbour hopping, the system shows infinite temperature MBL phase where all the many body states are localized. We then turn on either the power-law interactions keeping nearest-neighbour hopping or the power-law hopping keeping the nearest neighbour repulsion between fermions. Below we present results for various physical quantities analyzed, which clearly demarcate the effects of the long-range interactions from those of long-range hopping on the MBL phase.  All the results presented below are for the half-filled system for $h=5t_0$ and $V=t_0$, unless specifically mentioned.
\section{IV-A.  Level Spacing Statistics}
A powerful and basis independent measure of localization is based on the study of spectral statistics. The distribution of energy level spacings is expected to follow Poisson statistics (PS) for the MBL phase. The level spacing statistics of an ergodic phase is expected to be the same as that of a random matrix belonging to the same symmetry class. Hence, for the Hamiltonian in Eq.~\ref{model}, it should follow the Wigner-Dyson statistics (WDS).  We calculate the ratio of successive gaps in energy levels $r_n=\frac{min(\delta_n,\delta_{n+1})}{max(\delta_n,\delta_{n+1})}$ with $\delta_n=E_{n+1}-E_{n}$ at a given eigen energy $E_n$ of the Hamiltonian in Eq.~\ref{model}. For a Poissonian distribution, the disorder averaged value of $r$ is $2ln2-1\approx 0.386$; while for the Wigner surmise of Gaussian orthogonal ensemble (GOE) mean value of $r\approx 0.5295$.  

Fig.~\ref{rn} shows the level spacing ratio $\la r\ra$ averaged over the entire eigen spectrum for a  given disorder configuration and then averaged over many independent realizations of disorder. The energy resolved $\la r_n\ra$ vs normalized energy $\epsilon_n$ is shown in the appendix. For the system with power-law hopping (hollow squares in Fig.~\ref{rn}), $\la r\ra$ increases with decrease in $\alpha$. An increase in $\la r \ra$ with the system size is seen only for $\alpha <2$, with $\la r\ra$ approaching the value expected for Wigner surmise in the thermodynamic limit. This is a clear and strong signature of enhanced delocalization in the presence of long-range hopping. But for the system with power-law interactions (full circles in Fig.~\ref{rn}), $\la r \ra$ does not change significantly either with the change in $\alpha$ or the system size,  with $\la r\ra$ being always close to the PS value. This indicates that power-law interactions do not cause any delocalization of the system in consistency with the resonance argument described above and the quench dynamics discussed in the next section. 
\begin{figure}[h!]
\begin{center}
\includegraphics[width=2.2in,angle=-90]{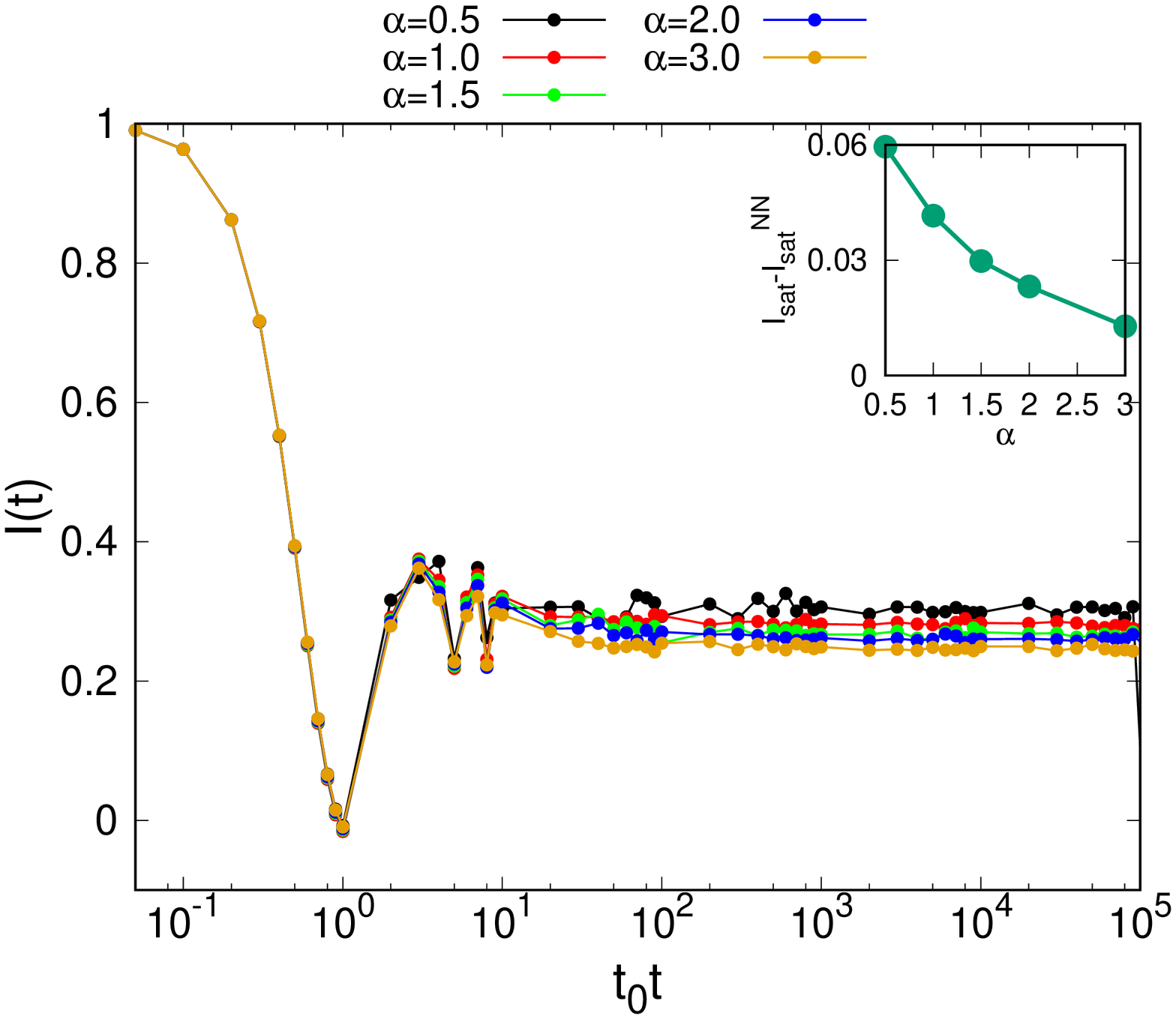}
\includegraphics[width=2.2in,angle=-90]{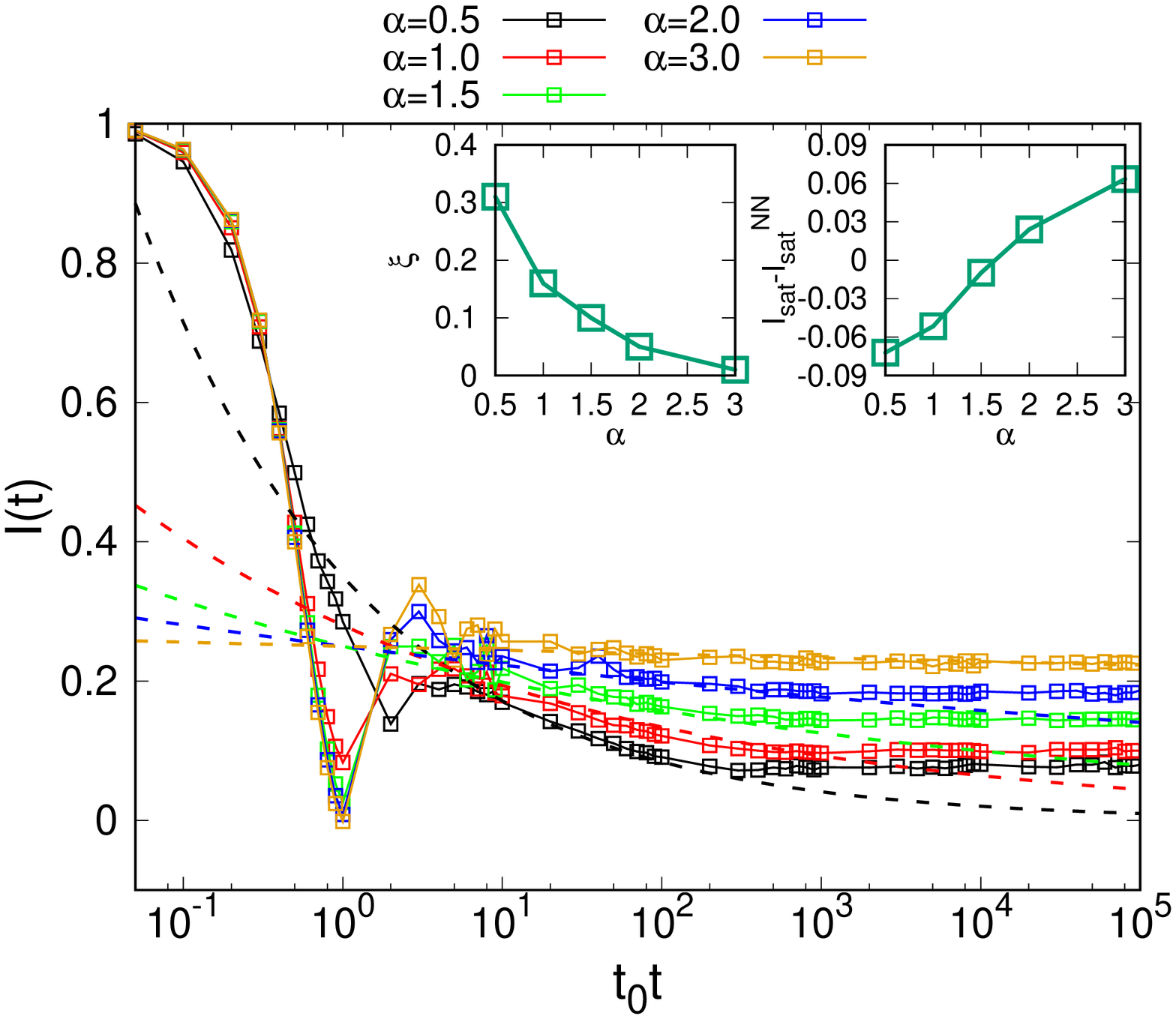}
\caption{The top panel shows the density imbalance $I(t)$ vs time for the system with power-law interactions for various powers of $\alpha$. For all values of $\alpha$, $I(t)$ saturates quickly after the initial rapid decay. The saturation value increases slightly for lower values of $\alpha$ as shown in the inset. There is no signature of delocalization due to long-range interactions. For the system with power-law hopping, shown in thr bottom panel, after the initial rapid decay  $I(t) \sim t^{-\xi}$ for intermediate time steps and then saturates. $\xi$ increases as $\alpha$ goes down and also the saturation value of $I_{sat}$ decreases with decrease in $\alpha$. This shows increase in the fraction of the delocalised states due to long-range hopping. The data shown is for $L=14$, obtained after averaging over 200 independent disorder configurations.} 
\label{imbalance}
\vskip-1cm
\end{center}
\end{figure}

\section{IV-B. Quench from a charge density wave: Density Imbalance and Entanglement growth}
While the number of resonances and the level spacing statistics are difficult to measure in experiment, there are other indications of MBL which can be experimentally probed. One of the most common method is monitoring the relaxation of an initial charge-density wave (CDW) order, in which all even sites are occupied~\cite{expt2}. To detect the localization properties of the system, the density imbalance between even and odd sites, $I(t) = \f{1}{L}\sum_i (-1)^i\la n_i(t)\ra$, is calculated as a function of time. Imbalance $I(t)$ is a signature of how much memory of the initial state the system has after certain time steps. Starting from an initial state $|\Psi_0\ra =\prod_{i=0}^{L/2-1} C^\dagger_{2i}|0\ra$ for the half-filled system, we let the state evolve w.r.t the Hamiltonian in Eq.~\ref{model} to obtain the time evolved state $|\Psi(t)\ra=exp(-iHt)|\psi_0 \ra$ and calculate the density imbalance as a function of time. In an MBL phase, $I(t)$, after a rapid initial decay saturates to a large constant value close to one due to strong memory of the initial state. On the other hand if some of the many-body states are delocalized, the system can relax and $I(t)$ decays as $I(t) \sim t^{-\xi}$ in the intermediate time scale after the initial rapid decay, before saturation in the long time limit. The saturation value decreases as the fraction of delocalized states in the system increases and is also a function of the system size.  

We calculated $I(t)$ numerically using exact diagonalization without making any approximation for the time evolution series. We basically write the initial state $|\Psi_0\ra$ as a linear combination of the exact eigenstates of the Hamiltonian in Eq.(\ref{model}), which gives $|\Psi_0\ra = \sum_n c_n |\Psi_n\ra$ where $H|\Psi_n\ra = E_n |\Psi_n\ra$. Using this we obtain the time dependent imbalance $I(t)=\f{1}{L}\la \Psi(t)|\sum_i (-1)^i n_i|\Psi(t)\ra = \f{1}{L}\sum_{n,m} c^\star_n c_m \la \Psi_n|\sum_i (-1)^i n_i|\Psi_m\ra \exp(i(E_n-E_m)t)$. 
The top panel of Fig.~\ref{imbalance} shows $I(t)$ for the case of long-range interactions. For all ranges of interactions studied $I(t)$ decays with almost the same rate and saturates to more or less the same value in the long time limit. No signature of delocalization is observed as $\alpha$ decreases, rather a close look (see inset) shows that the long time limit of $I(t)$ increases a bit as $\alpha$ decreases compared to its value for the system with nearest-neighbour repulsion. 

In contrast to this, in the presence of power-law hopping two significant changes in the relaxation of $I(t)$ are observed, specifically for $\alpha < 2$. For short-range hopping ($\alpha >2$), $I(t)$ shows saturation quickly after the initial decay just like in the case of power-law interactions, indicating that all the many-body states are localized. But for long-range hopping ($\alpha <2$), $I(t)$ relaxes as $I(t) \sim t^{-\xi}$ for the intermediate time steps, before saturating to a constant value. The exponent $\xi$ is larger for longer-range hopping, as shown in one of the inset of the bottom panel of Fig.~\ref{imbalance}, indicating enhancement in the fraction of delocalized state in the many-body spectrum~\cite{sdsarma_imb}. Secondly, the saturation value of imbalance $I_{sat}$ decreases with $\alpha$ eventually becoming smaller than $I_{sat}^{NN}$ for $\alpha < 2$, where $I_{sat}^{NN}$ is the saturation value of imbalance for the system with nearest-neighbour hopping and nearest-neighbour repulsion of the same strength. This is also an evidence of increase in the fraction of delocalized states due to long-range hopping. We will see clear signatures of it in the participation ratio in the Fock space and the energy resolved level spacing statistics as well. 
\begin{figure}[h!]
\begin{center}
\hspace{1cm}
\vskip-0.5cm
\includegraphics[width=2.3in,angle=-90]{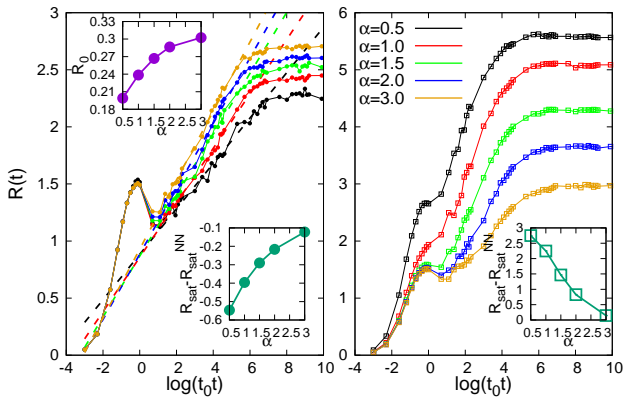}
\caption{The left panel shows the entanglement entropy as a function of time for the system with power-law interactions. The EE shows logarithmic growth $R \sim R_0 ln(t_0t)$ in time, a characteristic of the MBL phase , for all the values of $\alpha$. Both, the growth coefficient $R_0$ and the saturation value of the EE in the long time limit, decrease for smaller values of $\alpha$. The right panel shows the entanglement growth for the system with power-law hopping. Rate of growth of EE and the saturation value in the long time limit show monotonic increase as the range of hopping increases. This indicates that the system gets more delocalized in the presence of longer range hopping. The inset shows the long time value $R_{sat}$ in comparison to its value for the system with nearest neighbour hopping and nearest-neighbour interactions. The data shown is for $L=14$ averaged over 200 independent disorder configurations.}
\label{Renyi}
\vskip-0.5cm
\end{center}
\end{figure}

Furthermore, we study growth of entanglement entropy after a quench from the CDW initial state.  We evaluate the bipartite entanglement entropy (EE) by dividing the lattice into two subsystems A and B of sites $L/2$ and study the time evolution of the Renyi entropy $R(t) = -log[Tr_A \rho_A(t)^2]$ where $\rho_A(t)$ is the time evolved reduced density matrix obtained by integrating the total density matrix $\rho_{total}(t) = |\Psi(t)\ra \la \Psi(t)|$ over the degree of freedom of subsystem B. 
EE in an MBL phase is known to have logarithmic growth with time~\cite{Serbyn,logEE}. It has been established that the strongly disordered systems with short-range interactions show logarithmic growth of EE~\cite{Serbyn} with $R(t) \sim R_0 log(t_0t)$ in the strongly localized MBL phase, which can be explained in terms of de-phasing  due to exponentially small interaction induced corrections to the eigen energies of different states. The growth coefficient $R_0$ is known to be proportional to the single particle localization length. 
 Fig.~\ref{Renyi} shows $R(t)$ vs $log(t_0t)$ for various cases of power-law interactions. The dotted lines show the fit to $R_0log(t_0t)$ form with $R_0$ shown in the top inset. Even in the presence of long-range interactions, EE continues to show the logarithmic growth with time. The growth coefficient $R_0$ decreases as the range of interactions increases indicating even slower growth for longer-range interactions. Also the saturation value $R_{sat}$ in the long time limit increases with increase in $\alpha$. The bottom inset in the left panel shows $R_{sat}$ in comparison to its value for the system with nearest neighbour interactions and nearest neighbour hopping. 

In marked contrast to this, for the system with power-law hopping, growth of the EE increases with increase in the range of hopping. This also results in enhanced saturation value of the entanglement $R_{sat}$ for longer range hopping which indicates enhancement in the fraction of delocalised states. The EE in this case neither shows a clear logarithmic dependence on time, nor does it show a simple power-law form. For some time range the EE grows logarithmically with one slope and then the slope changes as $t_0t$ increases which results in a shoulder like structure of the EE after which the EE saturates. Note that for power-law hopping, EE has been shown to obey power-law behaviour with 
time~\cite{Tomasi,Ranjan} but only for large values of $\alpha \ge 3$, which is not the focus of interest in this work.

\begin{figure}[h!]
\begin{center}
\hskip-1cm
\includegraphics[width=2.3in,angle=-90]{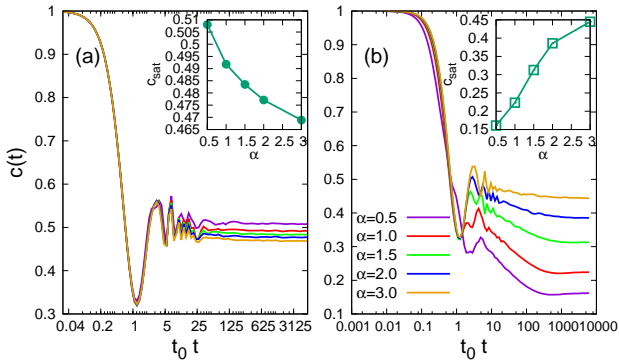}
\caption{Left Panel: Return probability $C(t)$ vs $t$ for the system with power-law interactions. $C(t)$ saturates quickly after the initial rapid decay, for all values of $\alpha$ studied, without showing any significant change as the range of interaction increases. Surprisingly the saturation value $C_{sat}$ decrease a bit for longer-range interactions. Right Panel: $C(t)$ vs $t$ for the system with power-law hopping. $C(t)$ decays as $t^{-n}$ after the initial rapid decay, the exponent $n$ being larger for the system with longer range hopping. This also results in smaller saturation values $C_{sat}$ indicating strong delocalization in the presence of long-range hopping. The data shown is for $L=16$ averaged over 200 configurations.}
\label{Ret_prob}
\end{center}
\end{figure}
\section{IV-C. Return Probability}
Another  dynamical quantity, which is a reliable measure of the memory of the initial state a system has, is the return probability. If there is a particle at site $i$ at time $t=0$, what is the probability that at a later time $t$ there is a particle at site $i$? We calculate the return probability defined as 
\be
C(i,t)=\frac{4}{V_f}\sum_{l=1}^{V_f}\langle l|(n_i(t=0)-\f{1}{2})(n_i(t)-\f{1}{2})|l\rangle
\ee

where $|l\rangle$ is the $l^{th}$ eigenstate of the Hamiltonian~(\ref{model}), and $n_i(t)$ is the density operator at site $i$ at time $t$. For a given disorder configuration, we average over all the sites followed by averaging over various independent disorder configurations to get the average value of the auto-correlation function $C(t)$. Return probability is a direct measure of the extent of localized states in the many-body spectrum~\cite{Subroto,garg}. For a fully localized system, $C(t)$ quickly saturates to a large value (close to one) after the initial rapid decay. For a system with a finite fraction of delocalized states, after the initial fast decay, $C(t)$ shows a power-law decay $C(t)\sim t^{-n}$ and then saturates. Larger value of the exponent, $n$ and the smaller saturation value of $C(t)$ is a signature of enhanced delocalization. 

The right panel of Fig.~\ref{Ret_prob} shows $C(t)$ for various values of $\alpha$ for the system with power-law hopping. As the range of hopping increases, $C(t)$ decays with a larger exponent $n$. Also, as shown in Fig.~\ref{Ret_prob}, the saturation value $C_{sat}$ is smaller for the system with longer range hopping indicating enhancement in the fraction of delocalized states. But in the presence of power-law interactions, $C_{sat}$ does not change significantly with increase in the range of interactions, though a small reduction in the saturation value $C_{sat}$ occurs for system with longer range of interactions. 
\begin{figure}[h!]
\begin{center}
\includegraphics[width=2.25in,angle=-90]{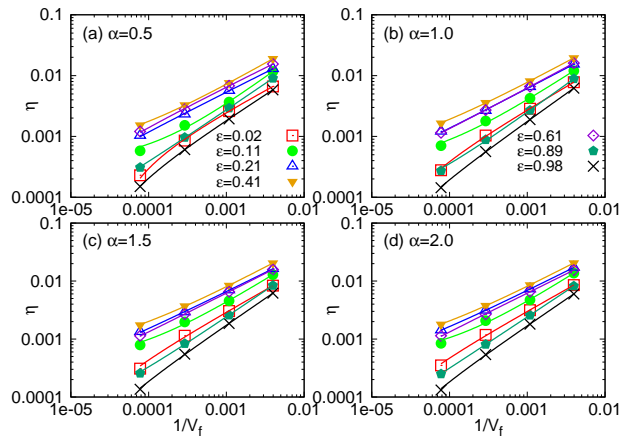}
\caption{$\eta(\epsilon)$ vs $1/V_f$, for a couple of $\epsilon$ values and for various values of $\alpha$ for the system with power-law interactions. No significant change in the scaling form is observed as the range of interactions increases. Almost all the many-body states remain localized even in the presence of long-range interactions for $\alpha <2$. The data shown is obtained by averaging over (4000-200) independent configurations for $L=10-16$.}
\label{eta_scal_int}
\end{center}
\end{figure}

\section{IV-D. Localization in the Fock Space}
\begin{figure}[h!]
\begin{center}
\hspace{-1cm}
\includegraphics[width=2.25in,angle=-90]{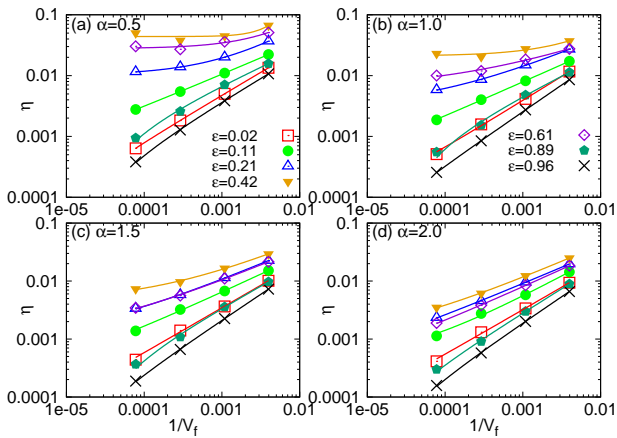}
\caption{$\eta(\epsilon)$  vs $1/V_f$ for a couple of $\epsilon$ values and for various values of $\alpha$ for the system with power-law hopping. As the range of hopping increases, more states in the middle of spectrum get delocalized. The data shown is obtained by averaging over (4000-200) independent configurations for $L=10-16$.}
\label{eta_scal_hop}
\end{center}
\end{figure}

In this section we analyze many-body wave functions in the Fock space.  To quantify the amount of localization in a given many-body state, we calculate the normalized participation ratio (NPR) which is defined as $\eta(\epsilon) = [\langle \sum_{i,n} |\Psi_n(i)|^4 \delta(E-E_n) \rangle_C V_f]^{-1}$, 
where $\Psi_n(i)$ is a normalized eigenfunction  with eigenvalue $E_n$ of the Hamiltonian in Eq.~\ref{model}, $V_f=$$^LC_N$  is the volume of the Fock space and $\langle \rangle _C$ represents the configuration averaging.  We use normalized energy $\epsilon =  \frac{E-E_{min}}{E_{max} - E_{min}}$ to show all our results.
$\eta(\epsilon)$ gives the fraction of configuration space participating in a many-body state of energy $\epsilon$.
For delocalised states $\eta(\epsilon) \sim a + b(1/V_f )^c$ while for the states showing MBL $a = 0$. This is clearly visible on the log scale plots shown in Fig.~\ref{eta_scal_int} and Fig.~\ref{eta_scal_hop} for $\eta(\epsilon)$ for various values of $\epsilon$.

\begin{figure}[h!]
\begin{center}
\includegraphics[width=2.35in,angle=-90]{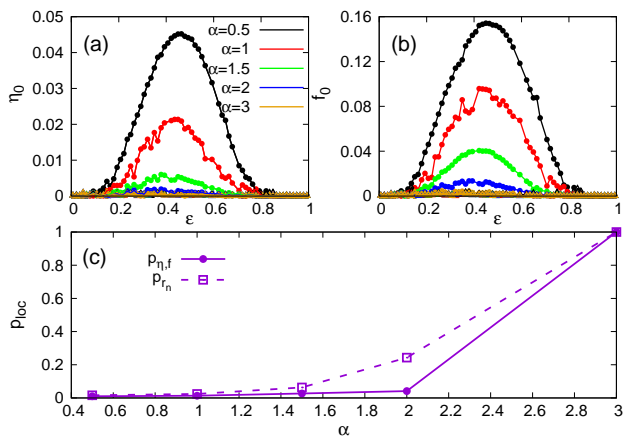}
\caption{ The top panel shows extrapolated values in the thermodynamic limit $\eta_0(\epsilon)$ and $f_0(\epsilon)$ vs $\epsilon$ for the system with power-law hopping. The range of energy over which $\eta_0,f_0 >0.001$ increases for smaller values of $\alpha$ which shows that more states get delocalized for longer range hopping. The bottom panel shows the fraction of localized states vs $\alpha$. $p_{r_n}$ is obtained from the energy resolved analysis of the level spacing ratio while the $p_{\eta,f}$ is obtained from the scaling analysis of NPR and Shannon entropy. Both indicate complete delocalization of the many-body spectrum for $\alpha <1 $ while for $1< \alpha <2$ a finite fraction of many-body states remain localized. For $\alpha >2$, infinite temperature MBL phase is observed.}
\label{eta0}
\end{center}
\end{figure}

Fig.~\ref{eta_scal_int} shows scaling of $\eta(\epsilon)$ for a couple of $\epsilon$ values for various ranges of interactions. For all the ranges of interactions studied $\eta(\epsilon)$ goes to a vanishingly small value in the thermodynamic limit, for all values of $\epsilon$, which shows that almost all the many-body states remain localized even in the presence of long-range interactions. There is no significant change in the scaling form or the extrapolated value of $\eta$ in the limit $V_f \rightarrow \infty$, with increase in the range of interactions. 
Fig.~\ref{eta_scal_hop} shows scaling plots of $\eta(\epsilon)$ for the system with power-law hopping. For $\alpha=2$, all the many-body states are localized with $\eta(\epsilon)\rightarrow 0$ in the thermodynamic limit. As $\alpha$ decreases below $2$, states in the middle of the spectrum get delocalized due to enhanced resonances mediated by long-range hopping. For $\alpha \le 1$, almost all the many-body states are delocalized. This is more clearly seen in the plot of extrapolated values $\eta_0(\epsilon)=lim_{V_f\rightarrow \infty} \eta(\epsilon)$ in Fig.~\ref{eta0}.

We have also evaluated Shannon entropy to distinguish between localized and delocalized states. For conventional, single particle Anderson localization problem, participation ratio is associated with the spread of a single particle state in the real space, but in the abstract Fock space there is no clear concept of distance between two points of the Fock space. Hence we used another measure to check for the localization in the Fock space, which is the Shannon entropy for every eigenstate  $S(E_n)=-\sum_{i=1}^{V_f}|\Psi_n(i)|^2\ln|\Psi_n(i)|^2$. Clearly for a many body state which gets contribution from all the basis states in the Fock space (and is normalized) $S(E_n)\sim \ln(V_f)$. Thus $f(\epsilon)=\exp(S(\epsilon))/V_f$ (obtained by averaging over all eigenstates in a small energy window around $\epsilon$) is of  order unity $f(\epsilon) \sim 1$ while for a localized state which gets significant contribution only from some of the basis states , say $N_l$, in the Fock space, $f(\epsilon) \sim N_l/V_f$ vanishes to zero in the thermodynamic limit. 
The scaling results for the Shannon entropy (not shown here) are consistent with those for the participation ratio, both indicating that the system shows enhanced delocalization in the presence of power-law hopping for $\alpha <2$, while no signature of delocalization is seen in the presence of power-law interactions for any value of $\alpha$ considered here.

 Fig.~\ref{eta0} shows the extrapolated values of NPR in the thermodynamic limit, $\eta_0(\epsilon)$, for the system with power-law hopping. As the range of hopping increases, not only the energy-range over which $\eta_0(\epsilon)$ is non-zero increases, but also the value of $\eta_0$ for the delocalized states increases. This shows that the fraction of delocalized states increases for longer range hopping. The right panel of Fig.~\ref{eta0} shows the extrapolated value of $lim_{V_f \rightarrow \infty}f(\epsilon) = f_0(\epsilon)$ in the thermodynamic limit which are consistent with $\eta_0$ plots and confirm that the fraction of delocalized states increases for longer range hopping.
 
The bottom panel of Fig.~\ref{eta0} shows the fraction of the many-body states that are localized, $p_{loc}$, as a function of $\alpha$ for the system with power-law hopping. The curve in full line is obtained from analysis of NPR and Shannon entropy while the dotted lines are obtained from the energy resolved level spacing ratio (shown in Appendix). The fraction of localized many-body states decreases for smaller $\alpha$, being vanishingly small for $\alpha \le 1$ while for $1< \alpha < 2$, a small fraction of the many-body states at the top and the bottom of the spectrum are localized. For $\alpha >2$ all the many-body states are localized. 

The analysis so far consistently illustrates that for a system in which all the single particle states are localized, and all the many-body states are localized too in the presence of nearest neighbour repulsion between fermions, turning on long-range interactions do not cause any delocalization of the many-body states though long-range hopping with $\alpha < 2$ causes delocalization. 
\begin{figure}[h!]
\begin{center}
\hspace{-1cm}
\includegraphics[width=2.35in,angle=-90]{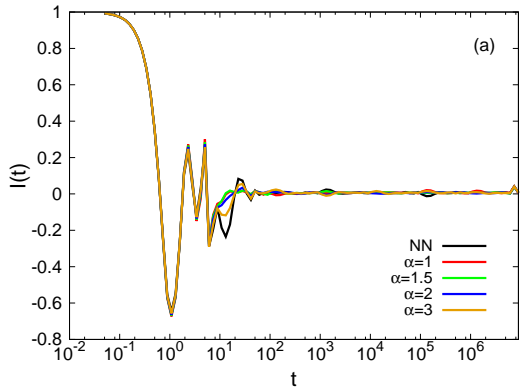}
\caption{Density imbalance for the AA model in the presence of power-law interactions. $I(t)$ decays to zero quickly for all values of $\alpha$ studied. This shows that all the many-body states remain extended in the presence of long-range interactions, provided that all the single particle states are extended in the non interacting limit. Results shown are for $h=0.5t_0$ and $V=0.3t_0$, obtained on $L=14$ size system, and averaged over 200 independent configurations.}
\label{AA}
\end{center}
\end{figure}

One might suspect, based on all these observations, that interactions themselves might be inducing localization in the system under consideration. To examine this possibility, we study Aubry-Andre model ($n=1$ case of the aperiodic model in Eq.~\ref{model}) in the presence of long-range interactions. We chose to work with $h=0.5t_0$ for which all the single particle states delocalized. To analyze the effect of power-law interactions on this system, we choose half filled system with $V=0.3t_0$. 
We did the quench analysis starting from a half filled CDW state and the result is presented in Fig.~\ref{AA}. For all the values of $\alpha$ studied, $I(t)$ quickly decays to zero which shows that the system remains to be fully ergodic and delocalized even in the presence of long-range interactions. This rules out the possibility of interaction induced localization in the presence of long-range interactions in this system.

At this point it is natural to ask the question, can long-range interaction delocalize a system which has single particle mobility edges in the non-interacting limit? We answer this question in the next section. 
\section{V. Effect of long-range interactions in the presence of single particle mobility edges}
In this section we study the model in Eq.~\ref{model} with parameter chosen such that the non interacting system has single particle mobility edges. We further chose $V$ such that the system with nearest neighbour interaction has a fraction of many-body states localized~\cite{garg}. But as shown in our earlier work~\cite{garg}, in order for the system to show MBL in this parameter regime, the system must be away from half filling.  Hence, we analyze the effect of power-law interactions on the quarter filled system. 
\begin{figure}[h!]
\begin{center}
\includegraphics[width=2.25in,angle=-90]{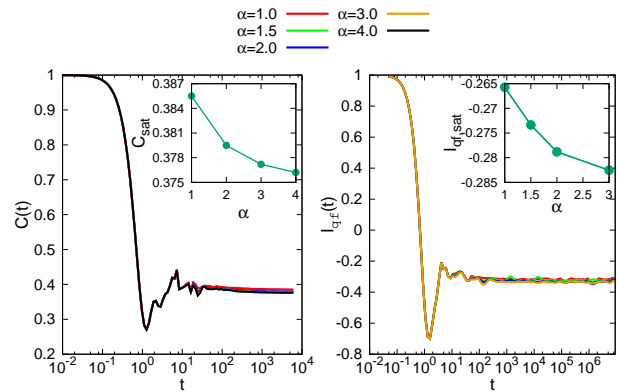}
\caption{ Left panel: Return probability $C(t)$ vs $t_0t$ for the system which has single particle mobility edges in the non-interacting limit. Data is shown in the presence of power-law interactions for various values of $\alpha$. No change is seen in the return probability as the range of interaction increases. The right panel shows the data for $I_{qf}(t)$ vs $t_0t$ for the same system. Results shown are for quarter filled system with $h=1.75t_0$, $V_1=0.3t_0$ and the power $n=0.5$ in the aperiodic potential for $L=16$ size chain obtained by averaging over 200 independent configurations.}
\label{qf}
\end{center}
\end{figure}

We did quench analysis starting from a charge density wave state at quarter filling, namely, $|\Psi_0\ra = |0001000100010001\ra$. We let this state evolve w.r.t the Hamiltonian in Eq.~\ref{model} for $h=1.75t_0$, $V=0.3t_0$, $V_1=V$ and $t_1=0$. This is the system with power-law interactions. The density imbalance is now defined as 
\be
I_{qf}(t)=\f{\sum_{i=0}^{L/4-1} n_{4i}-\sum_{i}^{\prime}n_i}{\sum_{i=1}^{L}n_i}
\label{Im_qf}
\ee
Here $\sum_i^\prime$ represents sum over all sites which are not included in the first sum of the numerator of Eq.~\ref{Im_qf}. By construction, at $t=0$, $I_{qf}=1$.
In a fully ergodic and delocalized phase, where the system does not have any memory of the initial state, the long time value of $I_{qf}^{sat} = -1/2$ because each site will be occupied with an equal probability of $1/4$ in such a state. The right panel of Fig.~\ref{qf} shows the time evolution of the density imbalance $I_{qf}(t)$ vs $t_0t$ for various cases of the power-law interactions. There is no signature of delocalization in the imbalance. For all power-law cases studied, the rate at which $I_{qf}(t)$ decays and the saturation value in the long time limit are almost independent of $\alpha$. 

We also looked for the effect of power-law interactions on the return probability $C(t)$. As shown in the left panel of Fig~\ref{qf}, the long-range interactions have basically no effect on the dynamics of the system. The rate with which $C(t)$ decays does not increase with range of interactions, rather the saturation value $C_{sat}$ increases very slightly with decrease in $\alpha$. 
This shows that even in a system which has both localized and extended states in the non-interacting limit, turning on long-range interactions do not delocalize the system~\cite{notes}. 
\section{VI. Conclusions and Discussions}
In this work, we have demonstrated that long-range interactions and long-range hopping have very different effects on the many-body localized phase. Though MBL persists in the presence of long-range interactions, long-range hopping in marked contrast to this, delocalizes the system. To be specific, we analyzed a system of spinless fermions with a deterministic aperiodic potential in one-dimension in the presence of either power-law interactions with nearest neighbour hopping or power-law hopping with nearest neighbour repulsion. We calculated the number of resonances in the Fock space by mapping the model described above to an effective Anderson model on a complex graph in the Fock space. Resonance calculation up to third order in $t_0/W$ does not show any signature of delocalization in the presence of long-range interactions (with $\alpha <2$) while the number of resonances diverge in the presence of long-range hopping. This can be understood in terms of the matrix elements of the effective Anderson model. Though interactions among all the particles in the system modify the local energy of this model, it is the hopping term which provides connectivity between various configurations in the Fock space. The range of interactions has a very weak effect on the local energies of the effective model, while the longer range hopping in real space, enhances the connectivity of configurations in the Fock space significantly, resulting in enhanced number of resonances proliferating through the Fock space which eventually delocalizes the system. 
 
Though in this work we have considered a system of spinless fermions with aperiodic potential, we believe that the above conclusion, at the broad level, holds true even for systems with fully random disorder. There might be detailed differences, as expected, in the time dependence of dynamical properties in the two set of systems. For a system of spin-$1/2$ particles, our study indicates that MBL will persist in disordered spin systems in the presence of long-range spin preserving interactions though long-range spin flip interactions will delocalize the system. In fact in one of our earlier work~\cite{sudip}, we established the existence of MBL-delocalization transition in the quantum Sherrington-Kirkpatrick model of Ising-spin glass which has infinite range Ising interactions. 

Starting from a system in which all the single particle states are localized, we analyzed various dynamical and static quantities in the presence of long-range interactions and long-range hopping, separately.  We showed that even in the presence of long-range interactions $\alpha <2$, the system continues to have long time memory of the initial state. This is consistent with experiments which have claimed to observe an MBL-like phase in a system with long-range interactions~\cite{ions,diamond}. In contrast to this, the long-range hopping with $\alpha \le 1$ delocalizes the system completely while a large fraction of many-body states are delocalized even for $1<\alpha<2$. It is interesting to compare the rough estimate of the critical value of $\alpha$ in our work with that obtained for long-range random hopping~\cite{Khatami}. In~\cite{Khatami}, violation of ETH is observed for $\alpha \ge 1.2$.

We also study effect of long-range interactions on a system which has, both, extended and localised single particle states separated by a mobility edge in the non-interacting limit. For the choice of parameters, such that the system with nearest-neighbour interactions has a fraction of many-body states localized, the system remains localized even in the presence of long-range interactions. 
 
At this point, we would like to mention that delocalization observed in earlier numerical studies of disordered systems in the presence of power-law interactions for $0< \alpha <1$~\cite{Burin,Mirlin} is because of the fact that both the long-range hopping and the long-range interactions are simultaneously present in the system. Long-range interactions alone do not delocalize the MBL phase.

\section{Acknowledgements}
A.G. gratefully acknowledges Science and Engineering Research Board (SERB) of Department of Science and Technology (DST), India under grant No. CRG/2018/003269 for financial support.
\\
\\
{\it Note Added}- After first submission of our manuscript, we became aware of a related study ~\cite{Roy} using a self-consistent mean-field theory of many-body localization for spin-1/2 chain in the presence of random disorder.  

\section{Appendix}
In this appendix we give details of the energy resolved level spacing statistics which has been used to obtain the fraction of localised states shown in Fig.[9] of the main text.

\begin{figure}[h!]
\begin{center}
\includegraphics[width=2.5in,angle=-90]{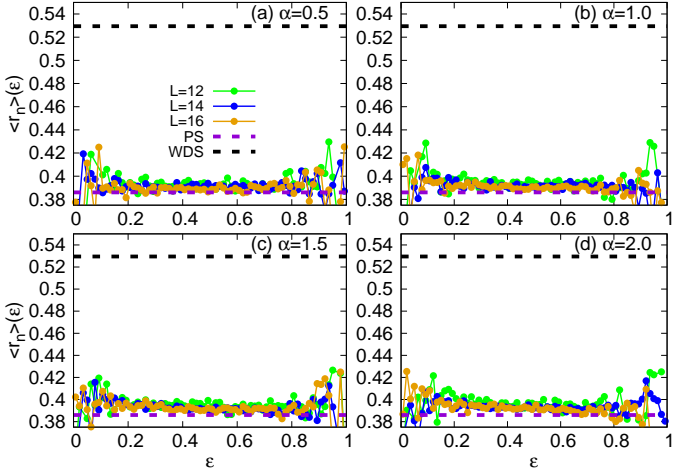}
\caption{Ratio of successive gaps $r(\epsilon)$ vs $\epsilon$ for the system with power-law interactions for various values of $\alpha$ and system sizes. For all values of $\alpha$ and $\epsilon$, $r(\epsilon)$ remains close to its value for the PS, and does not increase with the system size.}
\label{rint}
\end{center}
\end{figure}

Fig.~\ref{rint} shows the plot of $r(\epsilon)$ vs $\epsilon$ for the system with power-law interactions for various system sizes and for a couple of $\alpha$ values. This data is obtained from $r_n$ by averaging over (2000,1000 and 200) independent configurations for $L=12,14$ and $16$. For all the values of $\alpha$ studied, $r(\epsilon)$ remains close to the PS value for the entire spectrum of $\epsilon$. Also $r(\epsilon)$ does not show any significant change with the change in the system size. This shows that the range of interaction does not have any effect on the many-body localised system and the system persists to be in the MBL phase even in the presence of long-range interactions.
\begin{figure}[h!]
\begin{center}
\includegraphics[width=2.5in,angle=-90]{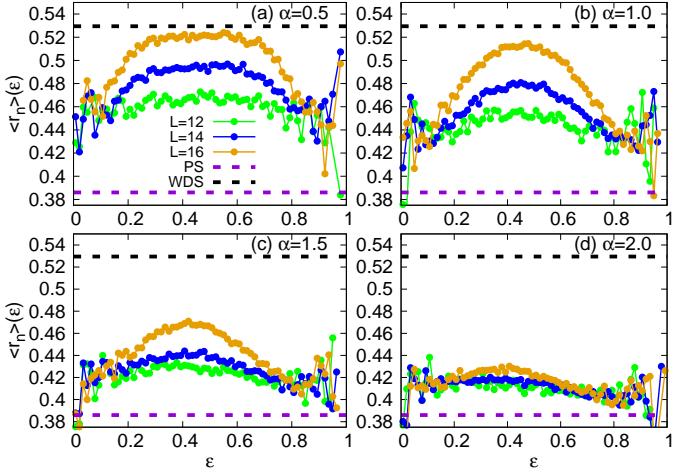}
\caption{Ratio of successive gaps $r(\epsilon)$ vs $\epsilon$ for the system with power-law hopping for various values of $\alpha$ and system sizes. For $\alpha=1.5$, many-body states on the top and the bottom of the spectrum show PS of $r(\epsilon)$ while for states in the middle $r(\epsilon)$ tends towards the WDS value as the system size increases. For $\alpha \le 1$, $r(\epsilon)$ approaches the WDS for large enough system sizes for almost all values of $\epsilon$.}
\label{rhop}
\end{center}
\end{figure}

Fig.~\ref{rhop} shows the plot of $r(\epsilon)$ vs $\epsilon$ for the system with power-law hopping. For $\alpha >2$, $r(\epsilon)$ remains close to the PS value for the entire spectrum and also does not show any significant increase with the system size, But for $\alpha < 2$, in contrast to the case of power-law interactions, here $r(\epsilon)$ shows PS value only for a small fraction of the many-body states on the top and the bottom of the spectrum. In fact for $\alpha  \le 1$, the entire spectrum shows $r(\epsilon)$ close to the WDS value, indicating the enhanced fraction of delocalized states due to longer range hopping.


\begin{thebibliography}{99}
\bibitem{note_lr}
Note that here we are not including discussion on confining potentials like $V(r)\sim r$ for which MBL phase has been claimed in ~\cite{Sondhi}.
\bibitem{notes}
There are two points to be noted here. Firstly, though a close inspection of the imbalance saturation value reveals a rather small decrease in $I_{qf,sat}$ as depicted in the inset and similarly a tiny increment of $C_{sat}$ for small values of $\alpha$, but within the small system size analysis it can not be considered as an indication of enhanced localization of the system. 
Secondly, both the imbalance and the return probability study indicate that long-range interactions have even weaker effects on this system with single particle mobility edges in comparison to the half-filled system presented in section IV. But this might be due to the fact that in order to have sufficient fraction of MBL states in the nearest neighbour model of repulsion, one has to choose to work at quarter-filling.
\end{thebibliography}
\end{document}